\documentclass[twocolumn,showpacs,preprintnumbers,amsmath,amssymb,floatfix]{revtex4}
\usepackage{graphicx}

\begin{document}

\title{Time-dependent universal conductance fluctuations in mesoscopic Au wires: implications}

\author{A. Trionfi, S. Lee, and D. Natelson}

\affiliation{Department of Physics and Astronomy, Rice University, 6100 Main St., Houston, TX 77005}

\date{\today}

\pacs{73.23.-b,73.50.-h,72.70.+m,73.20.Fz}

\begin{abstract}
In cold, mesoscopic conductors, two-level fluctuators lead to
time-dependent universal conductance fluctuations (TDUCF) manifested
as $1/f$ noise.  In Au nanowires, we measure the magnetic field
dependence of TDUCF, weak localization (WL), and magnetic field-driven
(MF) UCF before and after treatments that alter magnetic scattering
and passivate surface fluctuators.  Inconsistencies between
$L_{\phi}^{\rm WL}$ and $L_{\phi}^{\rm TDUCF}$ strongly suggest either
that the theory of these mesoscopic phenomena in weakly disordered,
highly pure Au is incomplete, or that the assumption that the TDUCF
frequency dependence remains $1/f$ to very high frequencies is
incorrect.  In the latter case, TDUCF in excess of $1/f$ expectations
may have implications for decoherence in solid-state qubits.
\end{abstract}

\maketitle

Two-level systems (TLS) are ubiquitous localized excitations in
disordered solids, and can profoundly affect
thermodynamic, dielectric, and acoustic
properties\cite{Esquinazibook}.  In mesoscale metals, scattering of
phase coherent conduction electrons by TLS results in time-dependent
(TD) universal conductance fluctuations (UCF)\cite{BirgePRL89}.
Because of the TLS distribution, TDUCF typically have a measured
$1/f$ frequency dependence.  The interplay of TLS and conduction
electrons may be relevant to correlated electronic
states\cite{RalphPhysRevL92,RalphPhysRevL94} and
dephasing\cite{ZawadowskiPRL99,ImryEuroPhysL99,AfoninPRB02}.
Interest has recently been renewed due to the importance of $1/f$
noise in limiting coherence in candidate solid-state
qubits\cite{PaladinoPRL02,HarlingenPhysRevB04,KuPhysRevL05,SchnirmanPhysRevL05,GalperinPRL06}.

%"at low temperatures" was removed since TLS exist at all temperatures. 

Electronic quantum interference produces other phenomena used to
investigate decoherence, including weak localization (WL)
magnetoresistance\cite{BergmannPhysRep84}, and UCF as a function of
magnetic field
(MFUCF)\cite{WebbPRL85,SkocpolPRL86,DebrayPRL89,MaillyJP92}.  Analysis
of WL and TDUCF as a function of magnetic field is expected to give
identical coherence lengths\cite{AleinerPRB02}, $L_{\phi}(T)$, if
electron-electron scattering is the only small-energy-transfer
process, as expected in clean normal metals at low temperatures.  Even
at temperatures where electron-phonon scattering is relevant, equality
between the WL and TDUCF-inferred coherence lengths is still expected.
The temperature at which electron-phonon scattering becomes important
is clearly visible in a log-log plot of coherence length versus
temperature.  As temperature is increased, the slope of this curve
will become more negative (from $-1/3$ to $-3/2$) indicating a
crossover from electron-electron dominated dephasing to
electron-phonon dephasing.  Comparisons between $L_{\phi}^{\rm WL}$
and $L_{\phi}^{\rm TDUCF}$ in AuPd have shown strong {\it
agreement}\cite{TrionfiPRB04}, while comparisons in clean, weakly
disordered Ag films and wires have shown an unexpected disagreement
below $\sim$ 10~K\cite{HoadleyPRB99,TrionfiPRB05}, when
electron-electron decoherence begins to dominate electron-phonon
scattering.

We have suggested\cite{TrionfiPRB05} that this apparent disagreement
results from an analysis based on an incorrect assessment that the
TDUCF are \textit{unsaturated} - that is, that TLS-induced conductance
changes within a coherent volume are much smaller than $e^{2}/h$.  The
saturated or unsaturated character of TDUCF depends on the microscopic
nature of the TLS, and determines which expression is used to infer
$L_{\phi}^{\rm TDUCF}$ from the field dependence of the
noise\cite{StonePRB89}.  Without detailed microscopic knowledge of the
TLS in a given material, one cannot know {\it a priori} whether the
TDUCF will be saturated or unsaturated.  Since the TLS are assumed to
have a broad distribution of energy splittings and relaxation times,
they likely also have a broad distribution of impacts on the
conductance.  The longer the coherence length, the more of the TLS
distribution is sampled within a single coherent volume.

Previously, saturation has been assessed by a simple consistency
check\cite{BirgePRB90}: How many decades of frequency would be
necessary for the integrated TDUCF $1/f$ noise power, $S_{G}\equiv
S_{V}/(R^{4}I^{2})$, to equal the variance, $\delta
G_{\textrm{MFUCF}}^{2}$, of the MFUCF?  Here $S_{V}$ is the measured
voltage noise power, $R$ is the sample resistance, and $I$ is the
measuring current.  If a required bandwidth far in excess of the
$\sim$ 20 decades reasonable for TLS\cite{FengBook91} is found (as it
has been in
Refs.~\cite{BirgePRB90,HoadleyPRB99,TrionfiPRB04,TrionfiPRB05}, for
example), this implies {\it unsaturated} TDUCF noise.

In this paper, we show that the assumption of unsaturated TDUCF noise
is inconsistent with WL data and systematic measurements based on
tuning paramagnetic impurity and TLS concentrations.  Either the
theory of these mesoscopic phenomena in pure, weakly disordered metals
is incomplete, or there is a flawed assumption in the consistency
check described above.  We suggest that the most likely flaw is that
the TLS ensemble has a power spectrum that deviates from the assumed,
extrapolated $1/f$ distribution.  Any excess fluctuations at high
frequencies may have implications for decoherence of solid state
qubits.  We compare $L_{\phi}^{\rm WL}$ and $L_{\phi}^{\rm TDUCF}$,
and $S_{G}$ and $\delta G_{\textrm{MFUCF}}^{2}$ in quasi-1D Au
nanowires, in two sets of experiments.  First, we tune $L_{\phi}^{\rm
WL}$ by systematically varying the concentration of paramagnetic
impurities at the Au interface in repeated measurements on a single
sample.  Second, we systematically modify the TLS distribution by
surface passivation of the Au via a self assembled monolayer (SAM) of
alkanethiol molecules.  Analysis of the data before and after these
modifications shows the apparent disagreement between $L_{\phi}^{\rm
WL}$ and $L_{\phi}^{\rm TDUCF}$ results from incorrectly fitting the
TDUCF versus magnetic field data using the unsaturated crossover
function.

\section{Fabrication and measurements}

All samples were patterned on undoped GaAs by electron beam
lithography.  High purity (99.9999\%) Au was deposited using an
electron beam evaporator.  Samples ranged from 60-80~nm in width and
were all roughly 15~nm thick.  Each current or voltage lead is
1~$\mu$m wide, and the leads are spaced 20~$\mu$m apart edge-to-edge.
There are a total of seven leads branching off from each wire.  An
anomalous paramagnetic impurity effect was seen while using Ti as an
adhesion layer.  We used this deliberately in some samples to lower
the coherence length via a (99.995\%) Ti adhesion layer of 1.5~nm.
All other samples were made with no adhesion layer.  Samples were
placed in a $^4$He cryostat and all measurements were performed
between 2 and 14 K using standard lock-in
techniques\cite{TrionfiPRB05}.  To limit additional averaging
associated to the drive current, TDUCF and MFUCF measurements were
always made at the same currents.  An ac five-terminal bridge
measurement\cite{ScofieldRevSciInst87,TrionfiPRB04} is employed for
TDUCF and MFUCF measurements.

\begin{table}
\caption{Sample parameters for the four reported samples.  Sample A is
the annealed sample with Ti adhesion layer.  Samples B-D are all SAM
treated without Ti.  The resistivities are given at 2 K both pre and
post treatment (annealing or SAM assembly).}
\begin{tabular}{c c c c c}
\hline \hline
Sample & $w$~[nm] & $t$~[nm] & $\rho_{\textrm{pre}}$~[$\Omega$m] & $\rho_{\textrm{post}}$~[$\Omega$m]  \\
\hline
A & 80 & 15 & 7.76$\times 10^{-8}$ & 6.03$\times 10^{-8}$ \\
B & 70 & 15 & 6.87$\times 10^{-8}$ & 6.32$\times 10^{-8}$ \\
C & 75 & 15 & 8.31$\times 10^{-8}$ & 9.26$\times 10^{-8}$ \\
D & 85 & 15 & 8.31$\times 10^{-8}$ & 9.26$\times 10^{-8}$ \\
\hline \hline
\end{tabular}
\label{tab:samppar}
\end{table}

The pertinent sample parameters are all given in Table
\ref{tab:samppar}.  The samples were all measured in the same manner
except for the post-annealing sample, A. Due to a failed lead, the
measurement scheme after annealing was done with 83 $\mu$m between the
voltage leads instead of 41 $\mu$m.  In order to fairly compare the
noise power before and after the annealing process, the length
difference of the sample needed to be accounted for.  As shown in 
Ref. \cite{FengPRL86}, the normalized noise power
$S_{\textrm{R}}/R^{2}\propto L_{\textrm{z}}^{-1}$.  In order to
correct the post-annealing noise power, the post-annealing values were
multiplied by 83/41.  With the parameters in the table, typical
two-segment lengths probed by the TDUCF and MFUCF measurements have
resistances of around 2.5~k$\Omega$.  In all samples the thermal length
($L_{T}\equiv \sqrt{\hbar D/k_{\rm B}T}$, where $D$ is the diffusion
constant) is much smaller than the inferred $L_{\phi}$ values.

Samples using the Ti adhesion layer were placed in the evacuated
sample space of the cryostat within 2 hours of metal deposition.
After finishing the measurements, the samples were allowed to anneal
at room temperature in ambient lab conditions for at least a week.
The measurements were then repeated. The pure Au samples to be treated
with a SAM were allowed to anneal at room temperature for a minimum of
a week before they were placed in the cryostat and measured.  In this
way the pure Au samples are allowed to anneal prior to {\it any}
measurements.  We have found that this initial annealing of pure Au
samples alone slightly reduces the resistivity relative to the
pre-annealing value, but induces no other changes; furthermore,
subsequent annealing produces minimal changes even on the timescale of
several days.  Changes seen after self-assembly of the SAM are
therefore due to the SAM, rather than simply letting the samples sit a
little longer.  The pure Au samples were then soaked in a 1 mM
solution of dodecanethiol (CH$_{3}$(CH$_{2}$)$_{11}$SH) in ethanol for
$\sim$ 48 hours, and returned to the cryostat to repeat all
measurements.

\begin{figure}[h!]
\begin{center}
\includegraphics[clip, width=8cm]{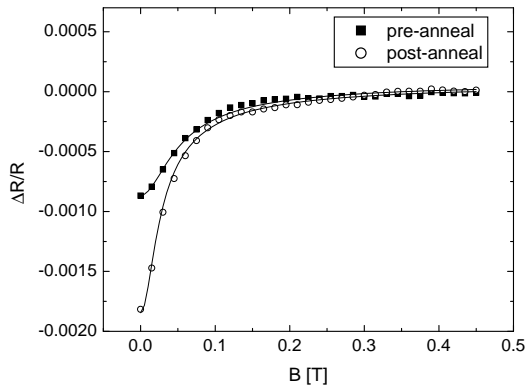}
\end{center}
\caption{The 2 K WL magnetoresistance of sample A before and after
annealing.  The size difference indicates a different coherence
length before and after annealing.  The solid lines are the
theoretical fit to the data with $L_{\phi}$ as the only fitting
parameter.} \label{fig:prbfig1}
\end{figure}

The WL magnetoresistance curves all showed strong
antilocalization, consistent with the large spin-orbit scattering
of Au.  Two magnetoresistance curves are shown in Figure
\ref{fig:prbfig1}.  The two curves are from sample A at 2 K before
and after annealing.  Magnetoresistance curves in the SAM treated
samples all looked similar in size and shape to the post-annealed
result of sample A.

Coherence lengths were inferred from the WL magnetoresistance
using\cite{PierrePRB03}:
\begin{eqnarray}
\frac{\Delta R}{R}\big{\vert}_{\rm 1d} & = & - \frac{e^{2}}{2 \pi
\hbar}\frac{R}{L} \times \nonumber \\
& & \hspace{-1.8cm} \left[3\left(\frac{1}{L_{\phi}^{2}}+\frac{4}{3
L_{\rm SO}^{2}}+\frac{1}{12}\left(\frac{w}{L_{\rm
B}^{2}}\right)^{2}\right)^{-1/2} \hspace{-0.6cm} -
\left(\frac{1}{L_{\phi}^{2}}+\frac{1}{12}\left(\frac{w}{L_{\rm
B}^{2}}\right)^{2}\right)^{-1/2}\right] \label{eq:1dwlso}
\end{eqnarray}
The value $\Delta R/R$ in this equation is defined as
$R(B)-R(B=\infty)/ R(B=\infty)$ while $L_{\rm SO}$ is the
spin-orbit scattering length, $w$ is the sample width, and $L_{\rm
B}\equiv \sqrt{\hbar/2eB}$. Both $L_{\phi}$ and $L_{\rm SO}$ are
left as free parameters while fitting.

\begin{figure}[h!]
\begin{center}
\includegraphics[clip, width=8cm]{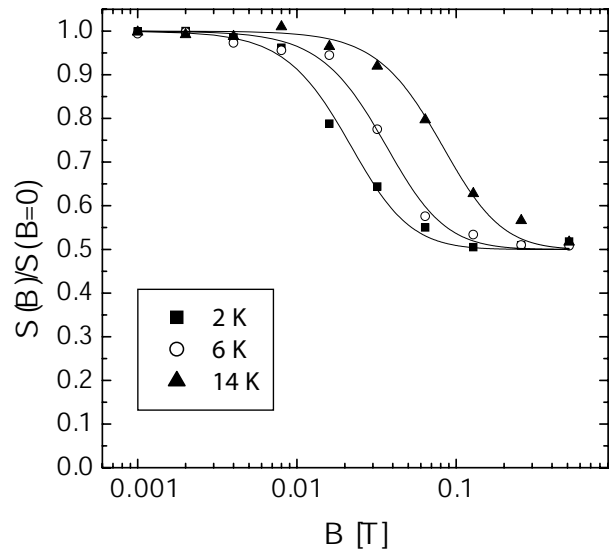}
\end{center}
\caption{The magnitude of the $1/f$ TDUCF noise as a function of
magnetic field for Sample C at three different temperatures, normalized
to its zero field value (see Eq.~(\ref{eq:xover}).  The
sample had been allowed to anneal at room temperature for one week
when this data was taken.  The solid lines are the theoretical fit to
the data assuming {\it unsaturated} TDUCF, with $L_{\phi}$ as the only
fitting parameter.} \label{fig:prbfig2}
\end{figure}

At each temperature the TDUCF are well described over the measured
frequency range by a $1/f$ frequency dependence of the noise power.
Examples of raw data for this in Sample A are shown below in
Fig.~\ref{fig:prbfig4}.  The coefficient of the $1/f$ dependence can be
measured as a function of magnetic field at each temperature.
Figure~\ref{fig:prbfig2} shows the typical field dependence for Sample
C.  As expected from theoretical considerations\cite{StonePRB89}, the
noise drops by a factor of two over a field scale that depends on the
coherence length, $L_{\phi}$.  The underlying physics is that the
breaking of time-reversal symmetry by the external field suppresses
the Cooperon contribution to the TDUCF, while the diffuson
contribution is unaffected.

Whether the TDUCF are saturated or unsaturated (as discussed above,
this depends on the detailed microscopic nature of the fluctuators)
determines the functional form used to infer $L_{\phi}$ quantitatively
from the data shown in Fig.~\ref{fig:prbfig2}.  When assuming {\it
unsaturated} TDUCF, we used an approximate crossover
function\cite{BeenakkerPRB88} of the form:
\begin{equation}
\nu(B) \equiv \frac{S_{G}(B)}{S_{G}(B=0)} = \frac{1}{2} + \frac{F'(B)}{2 F'(B=0)}.
\label{eq:xover}
\end{equation}
where
\begin{equation}
F'(B)=\frac{L_{\phi \rm{B}}^{5} \left(1 + \frac{3 L_{\phi
\rm{B}}^{2}(B)}{2 \pi L_{\rm T}^{2}}\right)}{4 \left(1 + \frac{9
L_{\phi \rm{B}}^{2}(B)}{2 \pi L_{\rm T}^{2}}\right)^{2}} +
\frac{3L_{\phi \rm{B} \rm{t}}^{5} \left(1 + \frac{3 L_{\phi \rm{B}
\rm{t}}^{2}(B)}{2 \pi L_{\rm T}^{2}}\right)}{4 \left(1 + \frac{9
L_{\phi \rm{B} \rm{t}}^{2}(B)}{2 \pi L_{\rm T}^{2}}\right)^{2}}
\label{eq:fderso}
\end{equation}
and
\begin{eqnarray}
L_{\phi \rm{B}}^{2}(B)& =& \frac{3 L_{\phi}^{2}}{(BeL_{\phi}
w/\hbar)^{2}+3},\nonumber \\
L_{\phi \rm{B} \rm{t}}^{2}(B)& = &\frac{3
L_{\phi}^{2}}{(BeL_{\phi}w/\hbar)^{2}+3 + 4
(L_{\phi}/L_{\rm{SO}})^{2}}.
\end{eqnarray}
The function $F'(B)$ is the derivative with respect to the coherence
time of the autocorrelation function of the magnetofingerprint, taken
when the TDUCF are unsaturated\cite{StonePRB89}.  To infer $L_{\phi}$
from the {\em saturated} crossover function, $F(B)$ is used instead of
the derivative.  Only $L_{\phi}$ was kept as a free parameter during
fitting, with $w$ and $L_{\rm SO}$ used from the WL fits.  Although
the saturated and unsaturated fitting functions give very different
coherence lengths when fit to TDUCF vs. B data, the graphical forms of
the two functions are {\it almost indistinguishable} by eye.  This makes it
difficult to determine whether a system is saturated or unsaturated
directly from TDUCF vs. B data.

The drive currents required to measure the TDUCF and its field
dependence are unfortunately much larger than those needed to measure
the WL magnetoresistance.  Concerns about Joule heating and adequate
thermal sinking of the electrons preclude extending the temperature
range of the TDUCF measurements down to dilution refrigerator
temperatures without some significant change in either sample
preparation or measurement technique.  

\section{Tunable magnetic impurity concentrations}

\begin{figure}[h!]
\begin{center}
\includegraphics[clip, width=8cm]{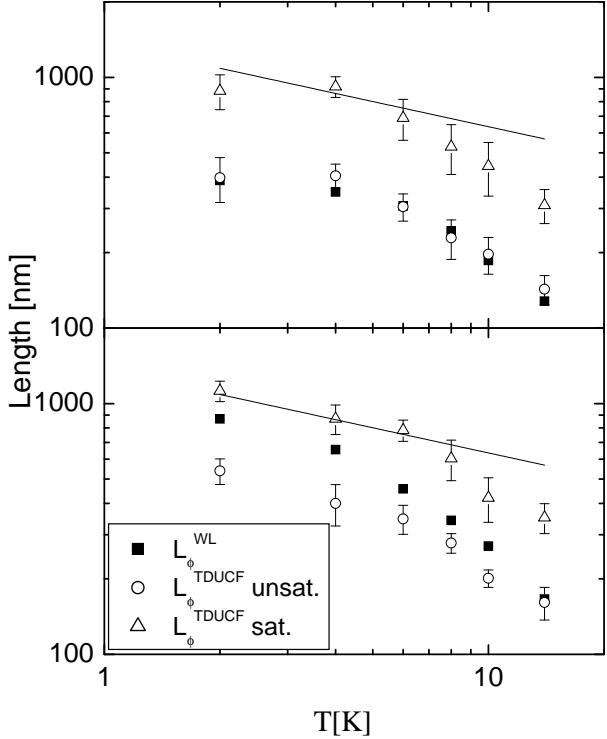}
\end{center}
\vspace{-3mm}
\caption{Coherence lengths inferred from both WL
magnetoresistance and TDUCF noise power versus magnetic field
before annealing (top graph) and after 2 weeks annealing (bottom
graph). The sample has a Ti adhesion layer of 1.5 nm. The solid
line is the theoretical Nyquist dephasing length.  } \label{fig:prbfig3}
\end{figure}

\begin{figure}[h!]
\begin{center}
\includegraphics[clip, width=8cm]{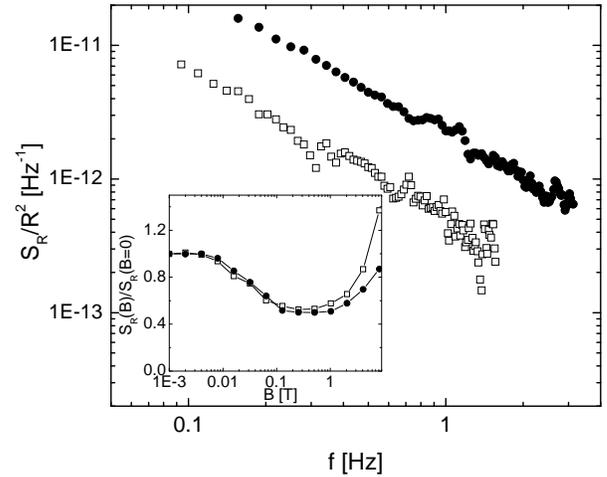}
\end{center}
\vspace{-3mm}
\caption{Normalized noise power vs. $f$ before(open) and
after (filled) sample annealing. The inset shows the noise
power as a function of magnetic field before and after annealing.
The larger upturn in the curve before annealing demonstrates a
larger paramagnetic impurity concentration.} \label{fig:prbfig4}
\end{figure}

Figure~\ref{fig:prbfig3} shows coherence lengths inferred from both WL and
TDUCF data in a sample with a Ti adhesion layer.  The data collected
before annealing show quite clearly that $L_{\phi}^{\rm TDUCF} \approx
L_{\phi}^{\rm WL}$ when unsaturated TDUCF are assumed.  Pre-annealing,
$L_{\phi}^{\rm WL}$ is much below the Nyquist length, consistent with
spin-flip scattering (from the Ti layer) as the dominant dephasing
mechanism at low temperatures.  This is reinforced by the inset in
Fig.~\ref{fig:prbfig4}, showing an upturn in noise power at high fields and
low temperatures attributed to Zeeman splitting of the paramagnetic
impurities.  After annealing in air, $L_{\phi}^{\rm WL}$ is much
increased, due to an apparent reduction in the paramagnetic impurity
concentration in the sample.  This is confirmed by the reduced size of
the upturn in the inset of Fig.~\ref{fig:prbfig4}, post-annealing.  As
we have discussed elsewhere\cite{TrionfiAPL06}, the paramagnetic
scattering sites are related to the oxygen stoichiometry of the
underlying adhesion layer, which is generally TiO$_{x}$, with $x \le 2$.

\begin{figure}[h!]
\begin{center}
\includegraphics[clip, width= 8cm]{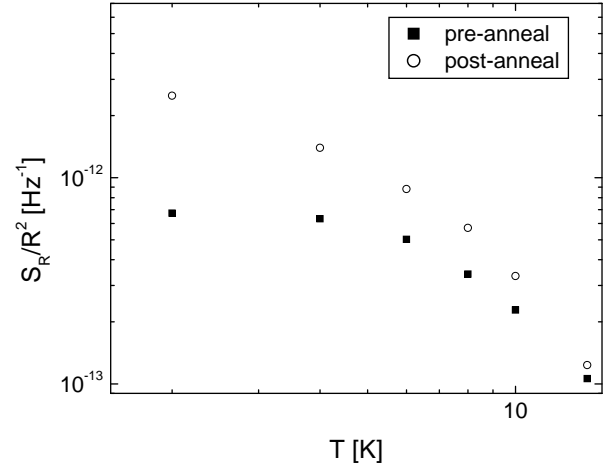}
\end{center}
\caption{The normalized zero-field noise power before and after annealing
with a 1.5 nm Ti adhesion layer.  The pre-anneal data is
consistent with a saturated coherence length by 2 K.}
\label{fig:prbfig5}
\end{figure}

The temperature dependence of the $B=0$ magnitude of $S_{\rm R}/R^{2}
\equiv S_{V}/(I^{2}R^{2})$  mirrors the $L_{\phi}^{\rm WL}$
data, as shown more clearly in Fig.~\ref{fig:prbfig5}.  Note the
saturation of noise power at low temperatures.  This indicates that
the coherence length is truly saturated (due to spin-flip scattering).
It is important to note that the noise power vs. temperature can be a
very subtle measurement.  Due to the signal-to-noise challenges in
measuring the $1/f$ resistance fluctuations, the noise power is
measured with a different drive current at each temperature.  Energy
averaging affects associated with the drive current
\cite{TrionfiPRB05} can suppress the {\it magnitude} of the $1/f$
noise without affecting the {\it normalized field-dependence} of the
$1/f$ noise.  This has also been demonstrated by Birge {\it et
al.}\cite{HoadleyPRB99}.  Because of this, comparing the magnitude of
the noise power at different temperatures should be done with care,
while the inferred coherence lengths (which depend instead on the
magnetic field dependence) are much more robust.  However, the
qualitative picture is still useful.  It should also be noted that
drive currents were unchanged pre and post treatment (i.e. the noise
power at 2~K was measured with the same drive current before and after
annealing).

The ability to tune the spin-flip scattering rate systematically in a
single sample through annealing allows us to see the effect of a
varying $L_{\phi}^{\rm WL}$.  After annealing, $L_{\phi}^{\rm WL}$ and
the unsaturated $L_{\phi}^{\rm TDUCF}$ (inferred from the TDUCF field
dependence) no longer agree below 14 K.  Such a disagreement was
reported previously\cite{HoadleyPRB99,TrionfiPRB05} in Ag, and we
suggested\cite{TrionfiPRB05} that this was due to a crossover from
unsaturated to saturated TDUCF with decreasing temperature (and
correspondingly increasing $L_{\phi}$).  In Fig.~\ref{fig:prbfig3}, the
likely explanation is that the true $L_{\phi}^{\rm
TDUCF}=L_{\phi}^{\rm WL}$ increased with annealing, pushing the TDUCF
farther into the saturated regime and rendering invalid the values
obtained from the unsaturated crossover function.  The unlikely
alternative is that the coherence physics without spin-flip scattering
affects WL and TDUCF differently, but large spin-flip scattering
washes out this difference.

The former interpretation is further supported by the data in
Fig.~\ref{fig:prbfig5}, as well as that in Fig.~\ref{fig:prbfig4} which shows
the normalized resistance noise power, $S_{\rm R}/R^{2}$, measured at
2~K before and after annealing.  The data have been normalized to
account for a change in lead configuration after annealing.  Clearly
the post-annealing noise is much larger.  This increase cannot be
accounted for by changes in the resistivity (post-anneal resistivity
is less than pre-annealing by roughly 10\%) or $L_{T}$.

There are only two possible explanations for this increase in noise.  We could
accept the unsaturated values of $L_{\phi}^{\rm TDUCF}$ in
Figure~\ref{fig:prbfig3} both pre- and post-annealing (the unlikely scenario
above), in which case the larger noise implies a factor of four {\it
increase} in the TLS concentration in the sample upon annealing.  This
is unreasonable, particularly in light of the decreased resistivity
after annealing attributed to increasing the grain size of the Au.
The more likely possibility is that annealing may have lowered the TLS
concentration while simultaneously increasing the true $L_{\phi}^{\rm
TDUCF}$.  The increased TDUCF amplitude then results from reduced
ensemble averaging as $L/L_{\phi}^{\rm TDUCF}$ decreases.  Another
observation that supports this idea is that the unsaturated
$L/L_{\phi}^{\rm TDUCF}$ in post-annealed samples becomes a closer
match to $L/L_{\phi}^{\rm WL}$ as the temperature is increased.  Much
like the increased spin-flip scattering lowered the coherence length,
as electron-phonon scattering begins to contribute to dephasing, the
coherence length of the system becomes much smaller, which would lead
to a sample further into the unsaturated TDUCF regime.

\section{Surface passivation}

Having seen the results of systematically tuning $L_{\phi}^{\rm WL}$,
we consider the complementary experiment, leaving $L_{\phi}$ fixed and
tuning the TLS density.  We performed measurements on three pure Au
samples (B, C, D), both before and after assembly of dodecanethiol.
The idea behind this series of measurements is to use the
self-assembled alkane chains to restrict the movement of atoms on the
wire surfaces.  If some of the TLS are due to these surface atoms,
then one would expect this SAM to alter the TLS distribution
accordingly.  

It is important to be sure that the changes observed in these SAM
experiments are truly due to the SAM, and not just the result of
further annealing.  In the case of Au on GaAs, annealing can cause
both the grain size to increase as well as the Au to wet the GaAs
causing width and thickness changes to Au wire.  Therefore several
precautions have been taken.  First, prior to any measurement those
samples have been allowed to anneal at room temperature for at least
one week.  This has been observed in the past to be a point beyond
which further room temperature annealing has essentially no effect on
the resistivity.  Since the self-assembly process takes place over 48
hours, we have also compared with the effects of simply letting the
samples sit for that period of time in methanol rather than a SAM
solution.  The effects shown below {\it only} happen as a result of
SAM assembly, and are qualitatively and quantitatively consistent
across the three samples.  The WL measurements pre- and post-assembly
also provide a means to check against size changes to the wire.  At
2~K, the WL fits always indicated small ($< 10 \%$) changes in the wire
width upon annealing, with no systematic increase or decrease in size.

\begin{figure}[h!]
\begin{center}
\includegraphics[clip, width=8cm]{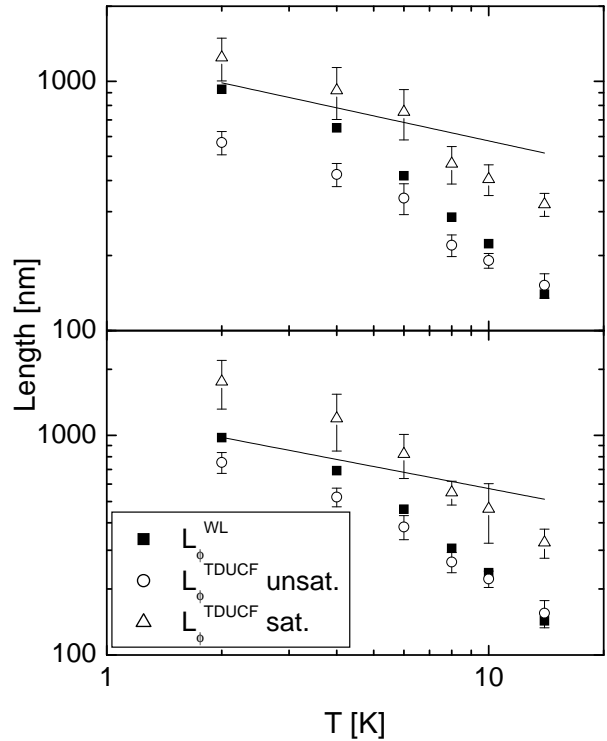}
\end{center}
\vspace{-3mm}
\caption{Coherence lengths inferred from WL and TDUCF noise power
versus magnetic field before (top graph) and after (bottom graph)
assembly of a dodecanethiol SAM.  The solid line is the theoretical
Nyquist dephasing time.  Solid squares are from weak localization
measurements.  Open circles assume unsaturated TDUCF, while open
triangles assume saturated TDUCF.} \label{fig:prbfig6}
\end{figure}

\begin{figure}[h!]
\begin{center}
\includegraphics[width=8cm, clip]{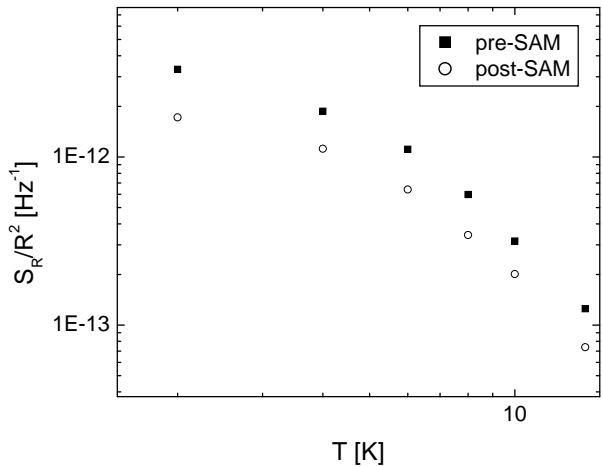}
\end{center}
\caption{The normalized noise power of sample B before and after
assembly of a dodecanethiol SAM. The sample has no Ti adhesion
layer.} \label{fig:prbfig7}
\end{figure}

Table~\ref{tab:samppar} shows that the self-assembly process has no
particular systematic effect on resistivity.  In two out of the three
samples, $\rho$ actually increases upon formation of the SAM.
Correcting for these slight changes in $\rho$, Figure~\ref{fig:prbfig6} shows
$L_{\phi}$ data in one such sample; all three showed similar results.
There was {\em no change} in $L_{\phi}^{\rm WL}$ due to SAM formation.
The noise power remained $1/f$ over the whole bandwidth, and its
measured magnitude {\it decreased} by a factor of $\sim 2$ over the
whole temperature range, with little change in the form of
the temperature dependence, as shown in Fig.~\ref{fig:prbfig7} for
the noise at zero field.

When the {\it field dependence} of the TDUCF is examined both before
and after self-assembly, there is an apparent increase in unsaturated
$L_{\phi}^{\rm TDUCF}$ due to the SAM.  That is, the field scale
over which the noise power is reduced by a factor of two as in
Fig.~\ref{fig:prbfig2} becomes {\it smaller}.  When the noise power
vs. field is fit using the unsaturated functional form of Eqs.~(\ref{eq:xover},\ref{eq:fderso}), the inferred $L_{\phi}^{\rm TDUCF}$ increases.
For example, the 2~K point shown in Fig.~\ref{fig:prbfig6} goes from
 $L_{\phi}^{\rm TDUCF} = 568$~nm before self-assembly to
 $L_{\phi}^{\rm TDUCF} = 753$~nm after self-assembly.  While
the error bars are not insignificant, this change exceeds the
error bar on the pre-SAM point by nearly a factor of three.

\begin{figure}[h!]
\begin{center}
\includegraphics[clip, width=8cm]{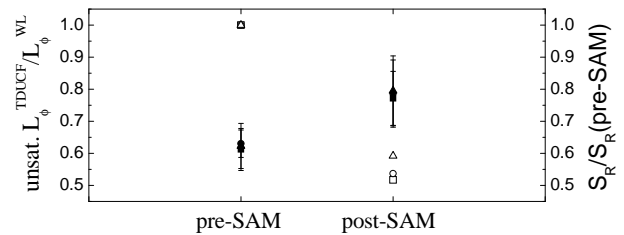}
\end{center}
\vspace{-3mm}
\caption{Filled shapes represent the ratios of the
unsaturated $L_{\phi}^{\rm TDUCF}$ to $L_{\phi}^{\rm WL}$ both
pre- and post-SAM assembly. The precision of the data coupled with
the large error bars indicate that the unsaturated crossover
function is not the correct functional form.  The open shapes show
the noise power ratios of pre- and post-SAM to pre-SAM assembly.}
\label{fig:prbfig8}
\end{figure}

This systematic change is seen in all three samples when
comparing pre- and post-SAM noise field dependence.
Figure~\ref{fig:prbfig8} shows the noise power at 2~K before
and after dodecanethiol exposure for all three samples tested, as
well as the ratio of the unsaturated $L_{\phi}^{\rm TDUCF}$ to the
$L_{\phi}^{\rm WL}$ at 2 K before and after the SAM assembly.

In order to accept the unsaturated $L_{\phi}^{\rm TDUCF}$ data as
correct (that is, as truly indicating an increase in coherence length
while the noise magnitude itself is reduced), the SAM would need to
simultaneously reduce the TLS concentration contributing to the TDUCF
as well as reduce some mysterious scattering rate that affects WL and
UCF differently.  We believe that the more likely explanation is that
as the SAM passivates TLS on the sample surface, the TDUCF move deeper
into the unsaturated regime and the unsaturated crossover function
becomes a better fit to the data.

The relatively large error bars on the coherence length ratios reflect
the unsaturated fitting function's systematic inability to thread all
the data points in the curve.  This inability can most likely be
attributed to the fact that the unsaturated fitting function is not
the correct functional form of the data being analyzed.  For
completeness, a similar comparison with the saturated $L_{\phi}^{\rm
TDUCF}$ resulted in the same qualitative situation of high precision
in the data points with large error bars.  A $\chi^{2}$ analysis
indicates similar ``goodness of fit'' for both unsaturated and
saturated functional forms of the field dependence.

\begin{figure}[h!]
\begin{center}
\includegraphics[clip, width=8cm]{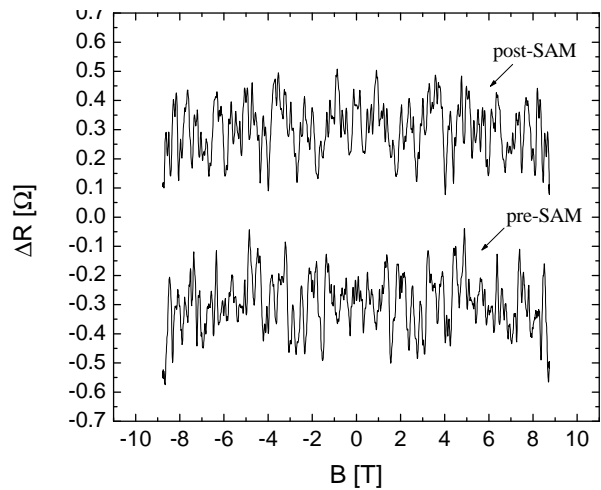}
\end{center}
\caption{The 2~K magnetofingerprint of sample B before and after SAM
assembly.
Only one sweep is shown for each curve.  The curves have been offset
for clarity.} \label{fig:prbfig9}
\end{figure}

For later comparison with the 2~K TDUCF data, we also measured MFUCF
at 2~K on these same samples.  Figure~\ref{fig:prbfig9} shows a
comparison of the MFUCF ``magnetofingerprint'' on sample B before and
after SAM assembly.  The WL magnetoresistance is eliminated by using
the 5 terminal measurement scheme.  Note the symmetry of the two
curves about zero; this demonstrates that the apparent noise is indeed
MFUCF.  Reproducibility of each curve was checked to confirm that the
fluctuations were actually a magnetofingerprint signature.

\section{Discussion}

We have seen in the two sets of experiments above that the coherence
lengths inferred from the TDUCF field dependence assuming {\it
unsaturated} TDUCF are very constraining.  In the Ti adhesion layer
case, the coherence lengths are initially relatively short due to
magnetic scattering from the adhesion layer.  In this limit
$L_{\phi}^{\rm WL}$ and $L_{\phi}^{\rm TDUCF}$ are in good agreement
with {\it no adjustable parameters}, similar to the results of
previous experiments on ``dirty'' samples with comparatively short
coherence lengths\cite{TrionfiPRB04}.  Annealing in air reduces
magnetic scattering, resulting in longer values of $L_{\phi}^{\rm WL}$
post-annealing.  This is reflected by an {\it increase} in noise power
magnitude, and a qualitative and quantitative change in the noise
power temperature dependence, all consistent with an increased
coherence length.  However, the field scale of the noise power
crossover is hardly changed.  Assuming unsaturated TDUCF, one then
finds that the inferred $L_{\phi}^{\rm TDUCF}$ {\it no longer agrees}
at all with $L_{\phi}^{\rm WL}$, even though the material is now {\it
cleaner}.

Similarly, SAM treatment reduces the TDUCF magnitude significantly, as
shown in Fig.~\ref{fig:prbfig7}, and $L_{\phi}^{\rm WL}$ is unchanged
after self-assembly, as is the temperature dependence of the noise
power.  However, there is a statistically significant {\it increase}
in $L_{\phi}^{\rm TDUCF}$ inferred from the noise field dependence
when unsaturated TDUCF are assumed.  Simultaneously increasing
$L_{\phi}^{\rm TDUCF}$ while decreasing noise magnitude is difficult
to understand from ensemble averaging considerations.

If the assumption of unsaturated TDUCF is what leads to this difficult
situation, it is important to check the validity of that assumption.
The MFUCF data shown in Fig.~\ref{fig:prbfig9} allow us to use the
approach of Birge {\it et al.}\cite{BirgePRB90} to check the
consistency of this assumption.  Before SAM exposure, $R =
2680.8~\Omega$, and after SAM assembly, $R = 2616.2~\Omega$.
Similarly, the variance in the MFUCF conductance at 2~K before the
assembly ${\rm var}~G_{\rm pre} = (1/R^{4}){\rm var}~R_{\rm pre} =
1.72 \times 10^{-16} \Omega^{-2}$.  After assembly, ${\rm var}~G_{\rm
post} = (1/R^{4}){\rm var}~R_{\rm post} = 1.71 \times 10^{-16}
\Omega^{-2}$.  Clearly the amplitude of the MFUCF is essentially
unaffected by the SAM, like $L_{\phi}^{\rm WL}$.

Converting from the normalized 2~K noise power plotted in
Fig.~\ref{fig:prbfig7}, $S_{\rm G}^{\rm pre} = \frac{S_{\rm R}^{\rm
pre}}{R^{4}} = 4.62 \times 10^{-19} \Omega^{-2}/{\rm Hz}$.  Similarly,
$S_{\rm G}^{\rm post} = \frac{S_{\rm R}^{\rm post}}{R^{4}} = 2.51
\times 10^{-19} \Omega^{-2}/{\rm Hz}$.  Assuming that the $1/f$
frequency dependence of the noise seen over our limited bandwidth
extends to much higher frequencies, as is commonly done, one can
estimate the necessary noise bandwidth if the TDUCF are saturated -
that is, the bandwidth required for the TDUCF contribution to be the
same magnitude as the MFUCF:
\begin{equation}
\log_{10}\frac{f_{\rm fin}}{f_{\rm in}} = \frac{{\rm var G}}{S_{G}}\log_{10} e.
\end{equation}
Plugging in, pre-SAM, $\log_{10}(f_{\rm fin}/f_{\rm in}) = 161.4$.
Post-SAM, $\log_{10}(f_{\rm fin}/f_{\rm in}) = 296$.  Since the
physically reasonable bandwidth of two-level systems ends at
frequencies comparable to the elastic scattering rate of the electrons
($\sim 10^{14}$~Hz), it is unphysical to think about 161 or 296
frequency decades of TDUCF.  Both of these are far in excess of the
physically reasonable 20 decades suggested\cite{BirgePRB90} as a rough
criterion of saturated TDUCF.  Therefore, in the conventional
analysis, one would conclude that the measured TDUCF are {\it
unsaturated}.

There are two clear possibilities: (a) The saturated/unsaturated
explanation of the data is somehow in error, requiring TDUCF and WL in
clean materials to be affected differently by common dephasing
mechanisms.  In other words the theory of these mesoscopic phenomena
in clean materials is incomplete. (b) Some assumption of the
consistency check is flawed.  We think that this is the more likely
possibility.  We typically measure the TDUCF noise spectrum up to a
few Hz.  Although the spectrum is $1/f$ between 100~mHz and 6~Hz in
these samples, and up to 100~Hz in other
work\cite{BirgePRL89,BirgePRB90,HoadleyPRB99}, the consistency check
assumes $1/f$ behavior to {\it arbitrarily high} frequencies.

A natural explanation for the failure of this consistency check would
be extra TLS spectral weight above the extrapolated $1/f$ magnitude at
higher frequencies.  Could such excess noise be detected?
Conservatively, suppose that the entire variance ${\rm var} G$ from
the MFUCF is made up by TDUCF that are {\it white} with respect to
frequency up to $\sim 10^{14}$~Hz.  This would be a worst-case
scenario for detectability.  An estimated white noise from these
excess fluctuations would then be $\sim {\rm var}G / (10^{14}~{\rm
Hz}) \approx 1.7 \times 10^{-30} \Omega^{-2}/{\rm Hz}$.  At a
measuring current pushing the limits of self-heating, this would
correspond to a voltage noise of $S_{V} = I^{2}R^{4}S_{\rm G, eff} =
1.8 \times 10^{-28}~{\rm V}^{2}/{\rm Hz}$.  This is approximately nine
orders of magnitude smaller than the Johnson noise from such a
resistor at 2~K.  Therefore, direct detection of the posited excess
noise would be unfeasible unless the fluctuators limit the excess
noise to a particular region of frequency space.

However, it is possible that this excess noise may be detectable at
lower temperatures and through its effects on other sensitive degrees
of freedom.  The possibility that the TLS-induced noise power has a
significant non-$1/f$ component at high frequencies has far-reaching
implications to the quantum computing community.  The internal noise
sources, i.e. TLS, can be the dominant dephasing mechanism in a qubit
when all other external mechanisms are filtered out
\cite{GalperinPRL06}. A non-$1/f$ noise power spectrum due to the TLS
found in normal metals could therefore result in an unexpected effect
on the dephasing of qubits.  Indeed, this may be the best way to probe
for further signatures of such noise.

We have performed two sets of experiments that examine the
relationship between $L_{\phi}^{\rm WL}$ and $L_{\phi}^{\rm TDUCF}$.
In some samples we have systematically reduced spin-flip scattering,
and find increased $L_{\phi}^{\rm WL}$, increased TDUCF magnitude, and
increased disagreement with $L_{\phi}^{\rm TDUCF}$ extracted assuming
unsaturated fluctuations.  In other samples we have passivated surface
fluctuators using a self-assembled monolayer, and find unchanged
$L_{\phi}^{\rm WL}$, decreased TDUCF magnitude, and better agreement
with unsaturated $L_{\phi}^{\rm TDUCF}$.  These results imply that
apparent disagreement between $L_{\phi}^{\rm WL}$ and $L_{\phi}^{\rm
TDUCF}$ likely results from a crossover from unsaturated toward
saturated fluctuations as $T \rightarrow 0$.  

On the one hand it is fortunate that such a crossover occurs in an
accessible temperature range for these experiments.  The currents
required for the TDUCF measurements and the resulting Ohmic heating
make it extremely difficult to extend these low frequency noise
measurements to dilution refrigerator temperatures.  On the other
hand, the fact that deviations between $L_{\phi}^{\rm WL}$ and
$L_{\phi}^{\rm TDUCF}$ have been observed in this temperature range
for almost fifteen years\cite{McConvillePRB93} was already an
indicator that interesting physics was taking place in the accessible
regime.  A simple comparison of integrated TDUCF and MFUCF magnitudes
fails to indicate such a crossover, suggesting that the assumptions
underlying that comparison are flawed.  We suggest that the
distribution of relaxation times for the TLS in Au may have extra
weight in excess of $1/f$ expectations at frequencies higher than the
measuring bandwidth of our experiments.  This extra high frequency
noise, should it exist, could have a strong impact on solid-state
qubits, and should be a focus of further research in electronic phase
coherence.

The authors thank N.O. Birge for his helpful advice concerning noise
measurements, and I.L. Aleiner and A.D. Stone for discussions of the
theory.  This work was supported by the David and Lucille Packard
Foundation and DOE Grant No. DE-FG03-01ER45946/A001.

%%%%%%%%%%%%%%%%%%%%%%%%%%

%\clearpage

%\clearpage

%\clearpage

\end{document}